\documentclass[prb,showpacs,twocolumn,amsmath,amssymb,floatfix]{revtex4-1}
\usepackage{array}
\usepackage{graphicx}
\usepackage{bm}
\usepackage[usenames]{color}
\usepackage{dsfont}
\usepackage{xspace}
\usepackage[applemac]{inputenc}

\newcommand{\beq}{\begin{equation}}
\newcommand{\eeq}{\end{equation}}
\newcommand{\beqa}{\begin{eqnarray}}
\newcommand{\eeqa}{\end{eqnarray}}
\newcommand{\re}{\mathrm{Re}}
\newcommand{\im}{\mathrm{Im}}
\newcommand{\hBN}{\textit{h}BN\xspace}
  
\begin{document} 
   
\title{Electronic structure of spontaneously strained graphene on hexagonal Boron Nitride}

\author{Pablo San-Jose, \'Angel Guti\'errez, Mauricio Sturla, Francisco Guinea}
\affiliation{Instituto de Ciencia de Materiales de Madrid, Consejo Superior de Investigaciones Cient\'ificas (ICMM-CSIC), Sor Juana In\'es de la Cruz 3, 28049 Madrid, Spain}

\date{\today} 

\begin{abstract}
Hexagonal Boron Nitride substrates have been shown to dramatically improve the electric properties of graphene. Recently, it has been observed that when the two honeycomb crystals are close to perfect alignment, strong lattice distortions develop  in graphene due to the moiré adhesion landscape. Simultaneously a gap opens at the Dirac point. Here we derive a simple low energy model for graphene carriers close to alignment with the substrate, taking into account spontaneous strains at equilibrium, pseudogauge fields and deformation potentials. We carry out a detailed characterisation of the modified band structure, gap, local and global density of states, and band topology in terms of physical parameters. We show that the overall electronic structure is strongly modified by the spontaneous strains.
\end{abstract}

\maketitle

\section{Introduction}

Graphene on hexagonal Boron Nitride (\hBN) was identified some years ago as a system with great promise for its remarkable electronic properties. \cite{Dean:NN10,Ponomarenko:NP11,Xue:NM11,Yankowitz:14,Kretinin:NL14,Tan:APL14} Graphene mobilities, when deposited on \hBN crystals, are dramatically improved, \cite{Dean:NN10,Gannett:APL11} as compared to those on SiO${}_2$ substrates. \cite{Kretinin:NL14,Tan:APL14} Similarly, charge inhomogeneities that obscure Dirac point physics are strongly suppressed by \hBN. Until recently, \hBN substrates have thus been conceived as an effective way to reveal the intrinsic electronic properties of graphene, remaining an otherwise effectively inert player. 

Recently, an interesting discovery has taken this point of view in a new direction. Graphene and \hBN are both hexagonal crystals, but have a lattice mismatch of around $\delta\approx 1.8\%$. It has been shown that when the two crystals are carefully aligned, a drastic structural and electronic reconstruction takes place. \cite{Woods:14} Driven by a competition between adhesion energy, which pushes graphene to deform locally in order to achieve commensuration with \hBN despite the mismatch, and elastic energy, which vies to keep graphene strain-free, a sizeable in-plane deformation pattern develops in graphene. Simultaneously, a spectral gap, measured at up to $\sim 30$ meV, opens at the Dirac point, which could prove to be particularly relevant for technological applications. These two features, elastic distortions and gap, disappear as graphene is rotated more than $\sim 1^\circ$ relative to \hBN. \cite{Jung:14,San-Jose:14}

Aligned graphene on \hBN has therefore emerged as the perfect playground to test the effects of strain on Dirac fermions. It is known \cite{Suzuura:PRB02,Neto:RMP09} that a finite strain field in graphene results in an effective pseudogauge field on low energy Dirac electrons, together with a scalar field, or deformation potential. Both are expected to have strong effects on electronic structure and transport, \cite{Guinea:NP09,Levy:S10} and even structural properties. \cite{San-Jose:PRL11}
A periodic strain superlattice, as the one developed by aligned graphene/\hBN, should therefore also be clearly visible in the electronic structure. The details, however, are not straightforward. As an example, neither the pseudogauge field nor the deformation potential will open a gap at the Dirac point by themselves (i.e. for electrically decoupled graphene and \hBN). The strain superlattice has the additional consequence of allowing the gap in \hBN to be slightly transferred onto graphene (in the form of an \hBN-induced self-energy), thereby making \hBN an electronically active component in the system. Other gap-opening mechanisms based on many-body interactions without strains have been studied \cite{Giovannetti:PRB07,Bokdam:PRB14, Jung:14}.

In this paper we present a low energy description of the electronic structure of graphene on \hBN, including the effects of an in-plane strain superlattice. Out of plane corrugations, explored in some recent works, \cite{Jung:14} are assumed negligible here based on experimental evidence. \cite{Yankowitz:NP12, Woods:14} Starting from the elasticity theory of Ref. \onlinecite{San-Jose:14} for spontaneous deformations in graphene-on-\hBN, we analytically characterise, in terms of physical system parameters, the associated corrections to the Dirac Hamiltonian of isolated graphene, including pseudogauge potential, deformation potential and \hBN-induced self-energy. We compute analytically the gap to third order in the corrections. We also present numerical results for the band structure, density of states (DOS), local density of states (LDOS), and band topology, for a collection of parameter choices. These results highlight the strong effect of spontaneous deformations on electronic structure, and can be used as a blueprint to experimentally identify the relevant effective model parameters for real samples, based on spectral and transport observations.

The paper is organised in six sections. In Sec. \ref{sec:Heff} we present the effective low energy Hamiltonian. In Sec. \ref{sec:Gap} we derive the associated gap in perturbation theory. In \ref{sec:DOS} we present numerical results for a range of spectral observables. In \ref{sec:Top} we describe the band topology of the effective model. Finally, in Sec. \ref{sec:Conclusion} we summarise our main results.

\section{Effective Hamiltonian \label{sec:Heff}}

Graphene on hexagonal Boron Nitride develops large distortions when both crystals approach perfect alignment. \cite{Yankowitz:NP12,Woods:14} These can be written in terms of a periodic displacement field $\vec u(\vec r)$ following the moiré pattern that results from the mismatch $\delta\approx1.8\%$ between the two crystals. \cite{San-Jose:14} The displacement field is symmetric under $2\pi/3$ rotations, with conjugate momenta $\vec G_{1,2}=\vec g_{1,2}\delta/(1+\delta)$, given in terms of graphene's conjugate momenta $\vec g_{1,2}$.  As was argued in Refs. \onlinecite{Woods:14,San-Jose:14}, the equilibrium displacements are the result of the competition between interlayer adhesion, which favours Carbon-on-Boron (AB) alignment, and graphene's elastic properties. The maximum local expansion $\frac{1}{2}\mathrm{Tr}\bm u=\frac{1}{2}(\partial_x u_x+\partial_y u_y)$ of graphene is reached at the center of AB regions, and approaches a value given by the lattice mismatch $\delta$. Local compression is concentrated in the other regions, and is of the same order $\delta$, but can also exceed this value depending on the strength of the adhesion. The equilibrium strain profile, evaluated within continuum elasticity theory, was derived in Ref. \onlinecite{San-Jose:14}. The solution remains smooth on the scale of graphene's lattice spacing, which results in a low energy electronic structure that preserves valley symmetry.

It has been argued \cite{Wallbank:PRB13,Mucha-Kruczynski:PRB13} that the simplest, yet generic, valley-symmetric model for the low energy electronic structure of a hexagonal graphene superlattice, as that created by an \hBN substrate, is given by the following first-star expansion
\begin{eqnarray}\label{H}
H&=& v \vec k\vec\sigma + w_0\sigma_0+\tilde{w}_3\sigma_3 \tau_3\nonumber\\
&&+\left[u_0 f_1(\vec r)+\tilde{u}_0 f_2(\vec r)\right]\sigma_0+
\left[u_3 f_2(\vec r)+\tilde{u}_3 f_1(\vec r)\right]\sigma_3 \tau_3\nonumber\\
&&+
\left(\hat z\times \frac{1}{|\vec G|}\left[u_\perp \vec\nabla f_2(\vec r)+\tilde{u}_\perp \vec\nabla f_1(\vec r)\right]\right)\vec \sigma \,\tau_3\nonumber\\
&&+\frac{1}{|\vec G|}\left[u_\parallel \vec\nabla f_2(\vec r)+\tilde{u}_\parallel \vec\nabla f_1(\vec r)\right]\vec \sigma\,\tau_3
\end{eqnarray}
Parameter $v$ is the Fermi velocity ($\hbar=1$), which may be expressed in terms of the effective nearest neighbour hopping amplitude $t$ and graphene lattice constant $a_0$ as $v=\sqrt{3}ta_0/2$. $\tau_3$ denotes the valley \footnote{A valley-isotropic basis is assumed. \cite{Akhmerov:PRL07}}.
The real functions $f_1(\vec r)$ and $f_2(\vec r)$ are even and odd combinations of first-star harmonics $\vec G_{1,2,3}$
\begin{eqnarray}
f_1(\vec r)&=&2\re\sum_{j=1}^3 e^{i\vec G_j\vec r}=2\sum_{j=1}^3 \cos{\vec G_j\vec r}\nonumber\\
f_2(\vec r)&=&2\re\sum_{j=1}^3 ie^{i\vec G_j\vec r}=-2\sum_{j=1}^3 \sin{\vec G_j\vec r} \label{f12}
\end{eqnarray}
where conjugate momentum $\vec G_3$ is defined as $\vec G_3=-(\vec G_1+\vec G_2)$ (it has the same modulus as $\vec G_{1,2}$). Parameters $w_0$ and $\tilde w_3$ represent a spatially uniform scalar shift and mass, respectively. The remaining corrections to the Dirac Hamiltonian are position dependent with zero average. The $u_\alpha$ terms  are even, and the $\tilde u_\alpha$ are odd under spatial inversion (note that $\sigma_{i\neq 0}$ and $f_2$ are odd). Physically, $u_3, \tilde u_3$ correspond to a position-dependent mass, while $u_\perp, \tilde u_\perp$ correspond to a (pseudo)gauge field, and $u_\parallel, \tilde u_\parallel$ to an (irrelevant) pure gauge, that is only included for completeness.
In principle, the values of the ten parameters $w_0$, $\tilde w_3$, $u_\alpha$ and $\tilde u_\alpha$ have to be derived from a microscopic model of the system. 

Such a microscopic derivation of the effective low energy Hamiltonian for graphene on an \hBN substrate is presented in Appendix \ref{sec:apHeff}. It is obtained, around each valley, by integrating out \hBN orbitals in the limit of a large \hBN gap $\Delta_{BN}$, and in the presence of the equilibrium distortion solution of Ref. \onlinecite{San-Jose:14}. The derivation yields, in principle, a Hamiltonian that contains harmonics beyond the first star. The extra harmonics, however, give very weak corrections to the electronic structure, and can be safely ignored, in favor of a simpler but still very accurate first-star model. Then, Eq. (\ref{H}) remains a good low energy description, given a proper choice of parameters. 
If we assume a negligible relative rotation between graphene ($\theta=0$), we may  write compact explicit forms for all Hamiltonian parameters,  in terms of the following quantities
\begin{eqnarray}
m_\pm&=&\frac{t_\perp^2}{2}(\epsilon_c^{-1}\pm\epsilon_v^{-1})\\
z_u&=&\frac{1}{9\sqrt{3}}\frac{1+\delta}{\delta^2}\frac{\epsilon_{AB}-\epsilon_{AA}}{B a_0^2}.
\end{eqnarray}
Here $\epsilon_{v,c}$ are the band edges of the valence (Nitrogen) and conduction (Boron) bands of \hBN, $t_\perp$ is the interlayer hopping, $B$ is graphene's bulk modulus, $a_0$ is its lattice constant, and $\epsilon_{AB}-\epsilon_{AA}$ is the adhesion energy per graphene unit cell of $AB$-stacked graphene on \hBN (Carbon-on-Boron), relative to that with $AA$ stacking (we approximate $\epsilon_{AA}\approx \epsilon_{BA}$ for simplicity, \cite{Giovannetti:PRB07,Sachs:PRB11,Jung:14}).
The effective parameters then read
\begin{eqnarray}
w_0&=&-\frac{1}{2}\left[\frac{1}{3}m_+ \left(15 z_u^2+2z_u+4\right)+m_-\left(z_u^2-2z_u\right)\right] \nonumber\\
\tilde{w}_3&=&\frac{1}{2}(m_+ + m_-)\left(z_u^2-2z_u\right)\nonumber\\
u_0&=&\frac{1}{12}\left[\frac{1}{3}m_+ \left(7 z_u^2-2z_u-2\right)-m_-\left(z_u^2+2z_u\right)\right]\nonumber\\
&&-\lambda_d z_u \delta/(1+\delta)\nonumber\\
\tilde{u}_0&=&\frac{1}{12\sqrt{3}}\left[m_+ \left(9 z_u^2+2z_u\right)-m_-\left(z_u^2+2z_u-2\right)\right]\nonumber\\
&&-\sqrt{3}\lambda_d z_u \delta/(1+\delta)\nonumber\\
u_3&=&\frac{1}{12\sqrt{3}}\left[m_+ \left(z_u^2+2z_u-2\right)+\frac{1}{3}m_-\left(z_u^2+2z_u\right)\right]\nonumber\\
\tilde{u}_3&=&\frac{1}{12}\left[m_+ \left(z_u^2+2z_u\right)-m_-\left(z_u^2+2z_u+2\right)\right]\nonumber\\
u_\perp&=&\frac{1}{18}\left[m_+\left(4z_u^2+z_u-2\right) + m_-\left(6z_u^2+3 z_u\right)\right]\nonumber\\
&&-\beta t z_u \delta/(1+\delta)\nonumber\\
\tilde{u}_\perp&=&\frac{1}{6\sqrt{3}}\left[m_+\left(6z_u^2+z_u\right)+m_- \left(4z_u^2-z_u-2\right)\right]\nonumber\\
&&+\sqrt{3}\beta t z_u \delta/(1+\delta)\nonumber\\
u_\parallel&=&\tilde u_\parallel=0
\label{usolution}
\end{eqnarray}
Parameter $z_u$ gives a dimensionless measure of the strain in the graphene layer [the maximum difference in local expansion \cite{San-Jose:14} is $\frac{1}{2}\Delta \mathrm{Tr}\bm u=[\partial_iu_i(\vec r_{AB})-\partial_iu_i(\vec r_{AA})]/2= 9 z_u \delta/(1+\delta)$], while $\beta\approx 2$ relates strains to pseudomagnetic fields in graphene, and $\lambda_d\sim 5 - 30$ eV  is the energy scale of the deformation potential (there is still quite some uncertainty as to its actual value \cite{Porezag:PRB95,Suzuura:PRB02,Pennington:PRB03,Bruzzone:APL11,Sule:JAP12}). For concreteness we use the following parameters: $z_u=-0.18$ (which corresponds to $\frac{1}{2}\Delta \mathrm{Tr}\bm u=2.8\%$, with $a_0=2.4$\AA,  $\delta=0.018$, $B=19.1$ eV/\AA${}^2$ and $\Delta\epsilon_{AB}=-100$ meV/unit cell), $t_\perp=0.3$ eV, $\lambda_d= 6$ eV, and $\beta=2$. Moreover we have taken $t=3.16$ eV, $\epsilon_c=3.34$ eV and $\epsilon_v=-1.4$ eV.\cite{Yankowitz:NP12} The resulting values for the effective parameters are summarised in Table \ref{tab:params}.
(We note that the results of Eq. (\ref{usolution}) and Table \ref{tab:params} neglect asymmetries between Carbon-Boron and Carbon-Nitrogen hopping amplitudes. Corrections that take this into account are discussed in Appendix \ref{sec:apHop}, but turn out to be parametrically small, and do not significantly affect the results that follow).

\begin{table}
   \centering
   \begin{tabular}{r|cccccccc} 
    & $w_0$ & $\tilde w_3$ & $u_0$ & $\tilde u_0$ & $u_3$ & $\tilde u_3$ & $u_\perp$ & $\tilde u_\perp$\\
   \hline
   $z_u=0$ &12	& 0	& 1	& 4.4	&1.8	&-7.6	&2.1	&-8.8 \\
   $\beta,\lambda_d=0$ & 3.9	& 5.3	& 2	&5.2	&-0.06	&-5.9	&1.3	&-7.5 \\
   $z_u,\beta,\lambda_d\neq 0$ & 3.9	& 5.3	& 21	& 38	& -0.06	&-5.9 	& 21	&-42  
   \end{tabular}
   \caption{Parameters of the effective Hamiltonian for representative physical parameters, in meV.}
   \label{tab:params}
\end{table}

\section{Spectral gap}
\label{sec:Gap}

The most prominent effect of the $\tilde w_3$, $u_\alpha$ and $\tilde u_\alpha$ perturbations on the Dirac Hamiltonian is to open a spectral gap at the (primary) Dirac point ($\Gamma$ point of the superlattice Brillouin zone). Another important effect is the development of secondary (possibly gapped) Dirac points away from neutrality at special locations along the boundary of the Brillouin zone. In this section we evaluate the magnitude of the primary gap, as a function of the model parameters.

To first order in perturbation theory, only the spatially uniform mass $\tilde{w}_3$ contributes to the gap, which reads simply $\Delta=2\tilde{w}_3\tau_3$ (we preserve the sign of the gap, that is opposite in each valley). Note that this contribution vanishes in the absence of strains ($\tilde w_3=0$ for $z_u=0$). To second order, we have a contribution to the gap from combinations of the position-dependent $u_\alpha$ terms, particularly from the combination $u_0 \tilde u_\perp+\tilde u_0 u_\perp$. These also vanish for $z_u=0$ (we have $u_0\tilde u_\perp=-\tilde u_0 u_\perp$ in that case). To third order, we have additional combinations involving $u_3$ and $\tilde u_3$ that, though small, remain finite even without strains (as long as $t_\perp\neq 0$). Adding up all contribution to third order, the gap can be written as
\begin{widetext}
\begin{eqnarray}
\label{gap}
\Delta=
2\tau_3&&\left\{\tilde{w}_3+
\frac{12}{v|\vec G|}(u_0\tilde u_\perp+\tilde u_0 u_\perp)\right.\\
&&\left.
+\frac{12}{v^2|\vec G|^2}\left[2\tilde{w}_3(u_3 u_\perp+\tilde u_3\tilde u_\perp)
-2(u_0u_3\tilde u_0+4u_3u_\perp\tilde u_\perp)-\tilde u_3(\tilde u_0^2-u_0^2+4u_\perp^2-4\tilde u_\perp^2-3u_3^2)-\tilde u_3^3\right]
\right\}+\mathcal{O}^4(u,w) \nonumber
\end{eqnarray}
where $v|\vec G|=2\pi t\delta/(1+\delta)$. Note that all combinations of parameters are odd under spatial inversion, so that operator $\Delta \sigma_3$ is even. Note also that the last term $\sim\tilde u_3^3$ was reported in Ref. \onlinecite{Song:PRL13a}, although it is typically weaker than the other contributions.
\end{widetext} 

\begin{figure*}
   \centering
     \begin{tabular}{ccc}
     \fbox{\includegraphics[width=0.44\textwidth]{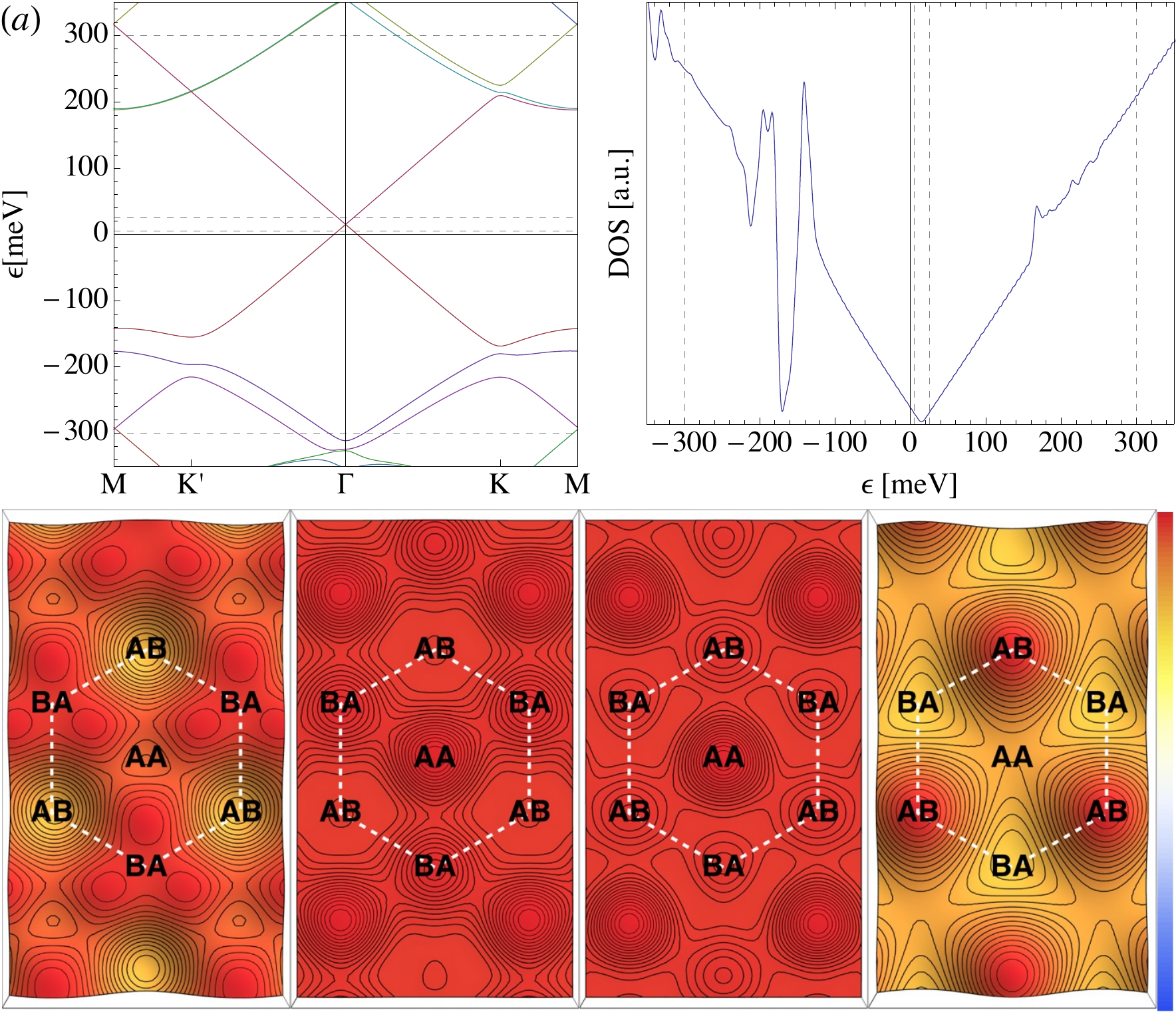}} & &
     \fbox{\includegraphics[width=0.44\textwidth]{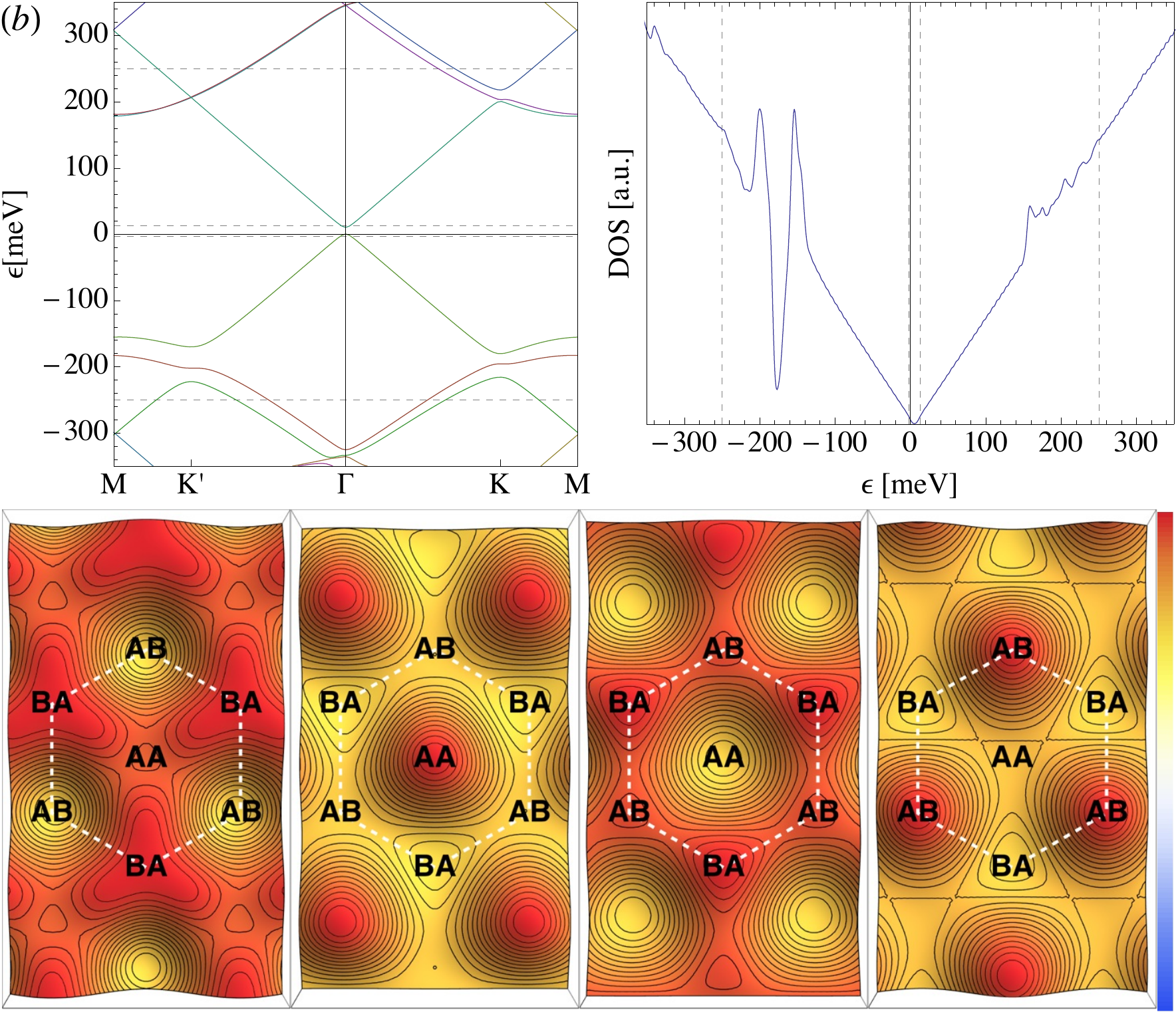}} \\
     $\left(z_u=0, t_\perp\neq 0, \lambda_d=0, \beta= 0\right)  \rightarrow \Delta\approx 0.14\,\mathrm{meV}$ & & 
     $\left(z_u\neq 0, t_\perp\neq 0, \lambda_d= 0, \beta=0\right) \rightarrow \Delta\approx 9.8\,\mathrm{meV}$
     \end{tabular}
     \caption{Comparison of the electronic structure for a model with pseudogauge field and deformation potentials switched off ($\beta=\lambda_d=0$), both with (right: $z_u\neq 0$) and without (left: $z_u=0$) strains. Represented are the low-energy band structure, the corresponding total DOS, and the LDOS in real space for four energies shown as dashed lines in the band structure and DOS. In the LDOS the coloring corresponds to blue for zero, and red for the maximum. We have included the value for the (non-perturbative) gap $\Delta$ at the primary Dirac point in each case. }
   \label{fig:ES1}
\end{figure*}

It is interesting to note that if we consider an expansion to leading orders of $z_u$ and $m_\pm$, instead of $\tilde w_3$, $u_\alpha$ and $\tilde u_\alpha$, both the first and second-order processes above contribute to the leading (second) order in the new expansion,
\begin{eqnarray}
\Delta=2\left\{(m_++m_-) z_u\left(\frac{2\lambda_d-\beta t}{\sqrt{3}\pi t}-1\right)\right\}+\mathcal{O}^3(m_\pm, z_u) \label{gappert2}
\end{eqnarray}
Note that the pseudogauge and the deformation potential contributions to the gap tend to cancel each other. This result neglects the adhesion energy difference $\epsilon_{BA}-\epsilon_{AA}$ (which is much smaller than $\epsilon_{AB}-\epsilon_{AA}$ \cite{Giovannetti:PRB07,Sachs:PRB11}), hence the particular dependence on $m_++m_-=t_\perp^2/\epsilon_c$, that includes only the conduction band edge of \hBN (Boron character). A more general perturbative expression, albeit with $\lambda_d=0$, can be found in Ref. \onlinecite{San-Jose:14}, but the difference is negligible in practice. 

\section{Electronic structure phenomenology}
\label{sec:DOS}

\begin{figure*}
   \centering
     \begin{tabular}{ccc}
     \fbox{\includegraphics[width=0.44\textwidth]{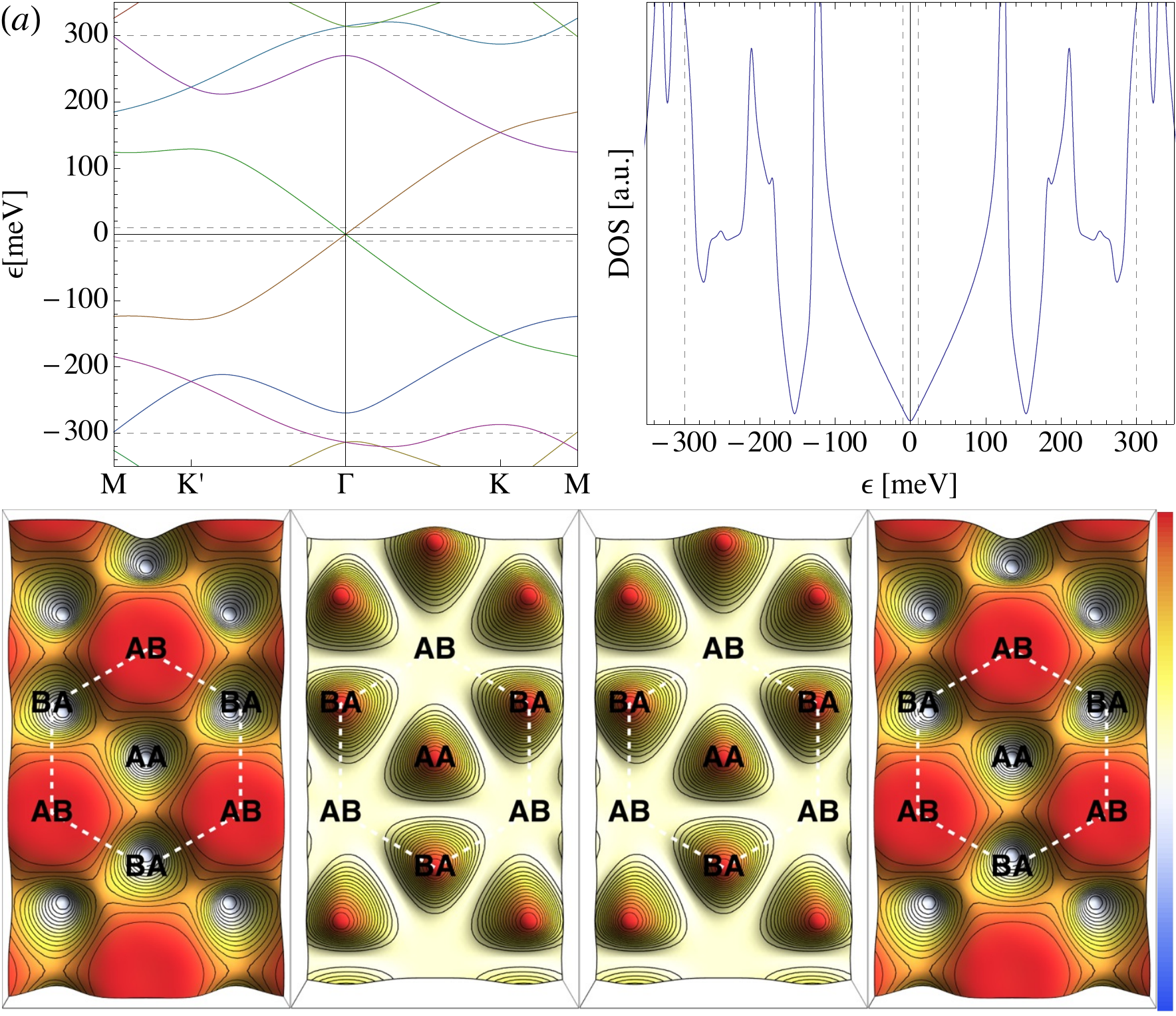}} & &
     \fbox{\includegraphics[width=0.44\textwidth]{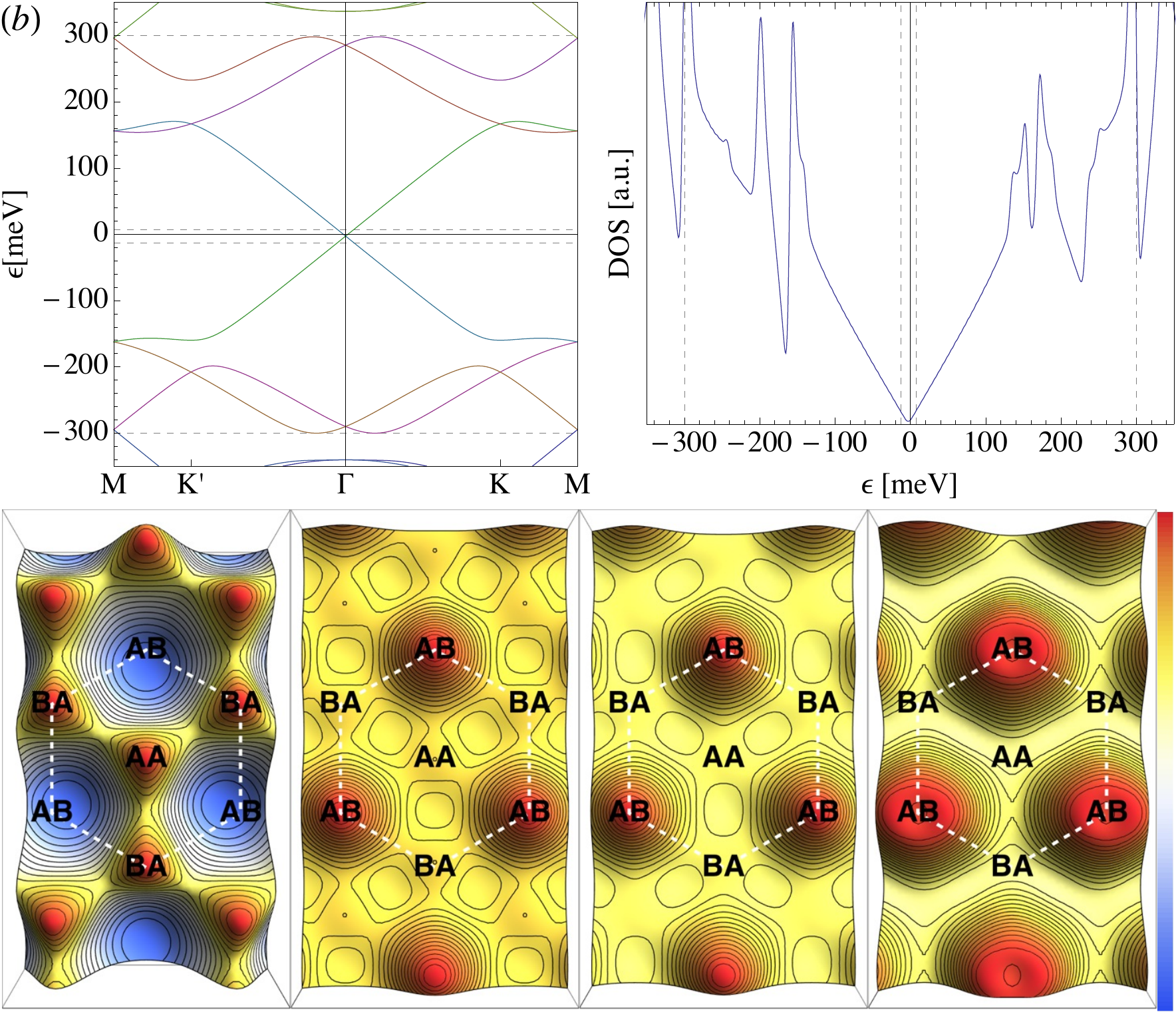}} \\
     $\left(z_u\neq0, t_\perp=0, \lambda_d=0, \beta\neq 0\right)  \rightarrow \Delta=0$ & & 
     $\left(z_u\neq0, t_\perp=0, \lambda_d\neq 0, \beta=0\right)  \rightarrow \Delta=0$
     \\ \\
     \fbox{\includegraphics[width=0.44\textwidth]{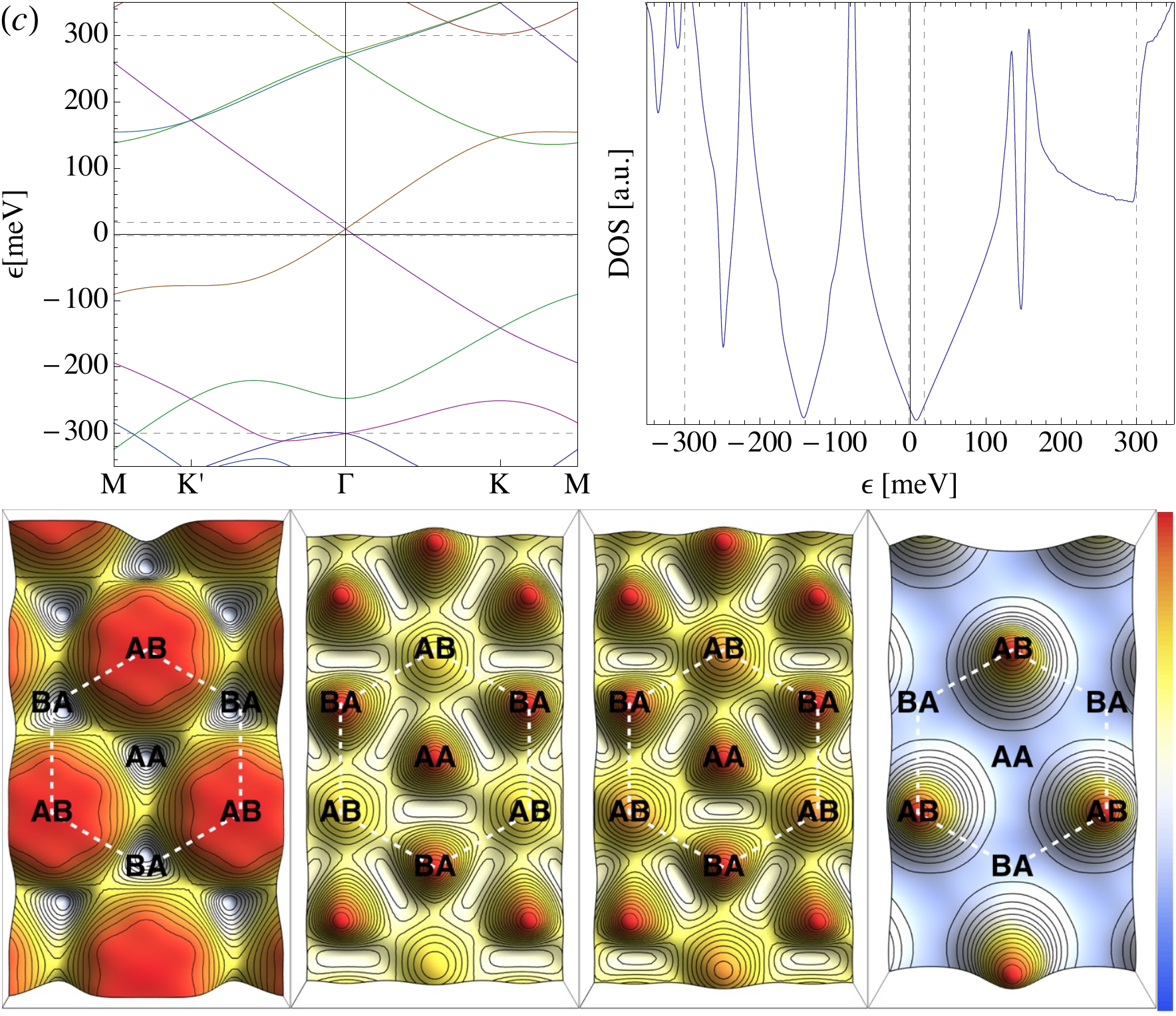}} & &
     \fbox{\includegraphics[width=0.44\textwidth]{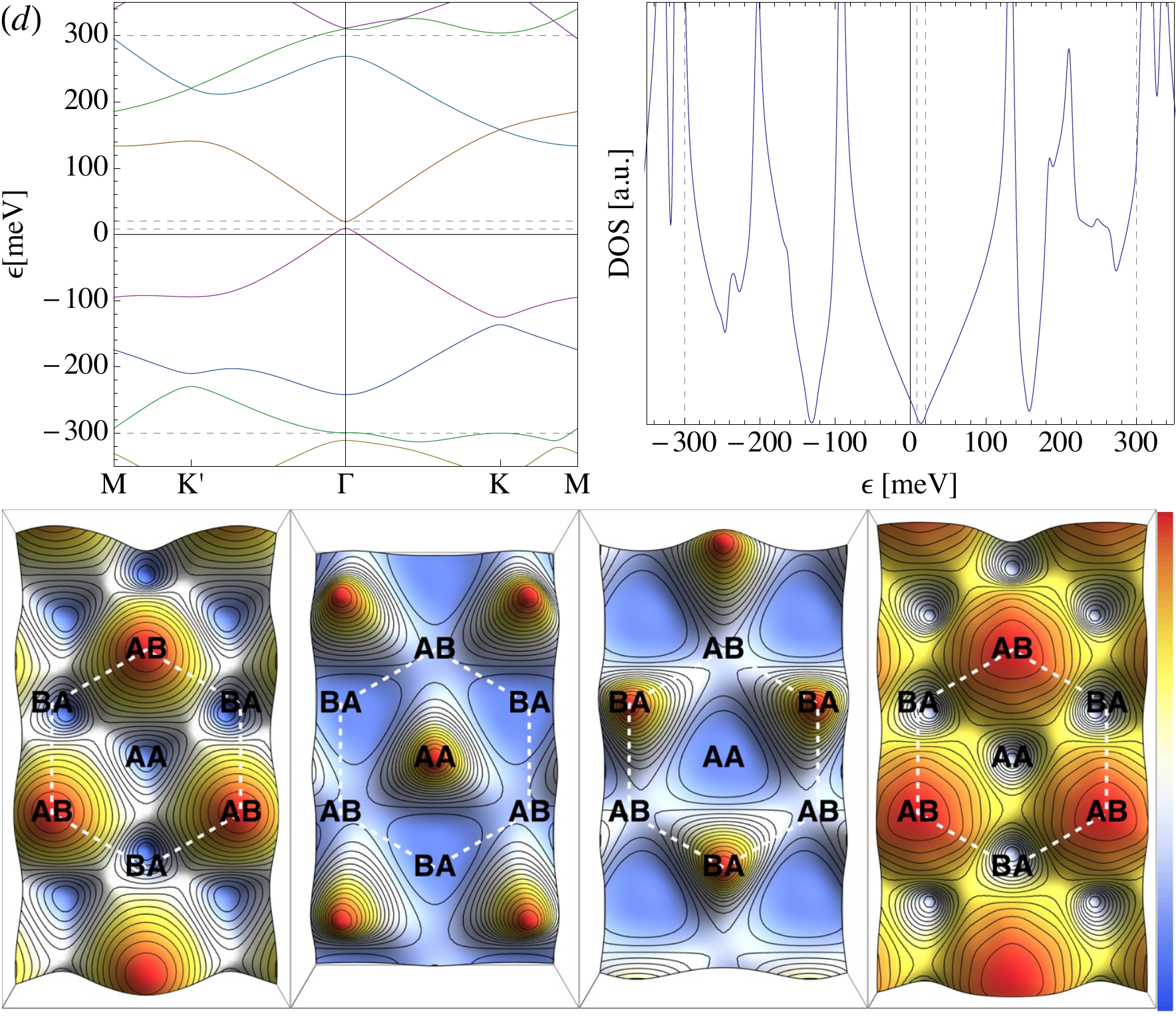}} \\ 
     $\left(z_u\neq0, t_\perp=0, \lambda_d\neq0, \beta\neq 0\right)  \rightarrow \Delta\approx 0\,\mathrm{meV}$ & & 
     $\left(z_u\neq0, t_\perp\neq0, \lambda_d= 0, \beta\neq 0\right)  \rightarrow \Delta\approx 9.6\,\mathrm{meV}$
     \\ \\
     \fbox{\includegraphics[width=0.44\textwidth]{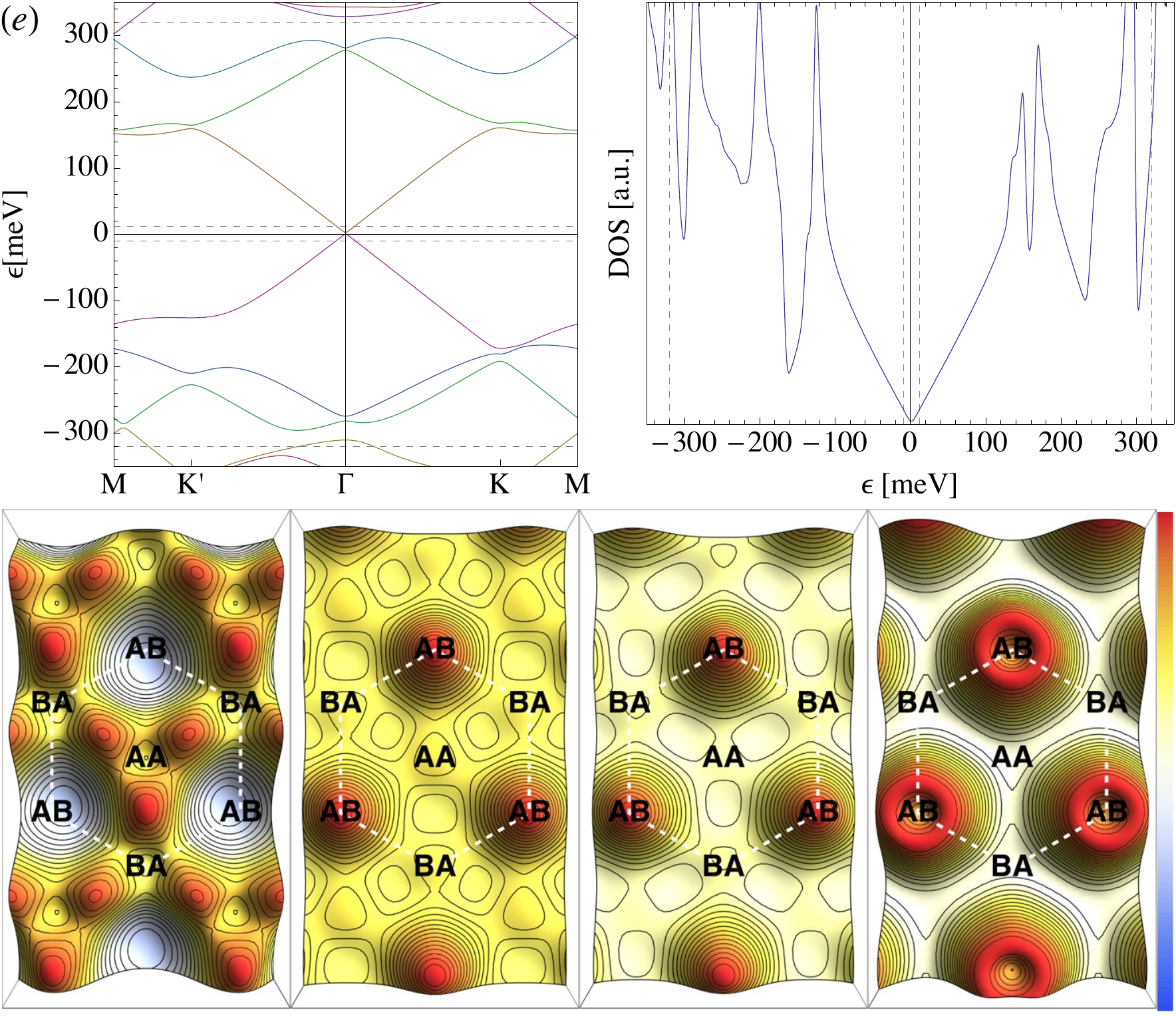}} & &
     \fbox{\includegraphics[width=0.44\textwidth]{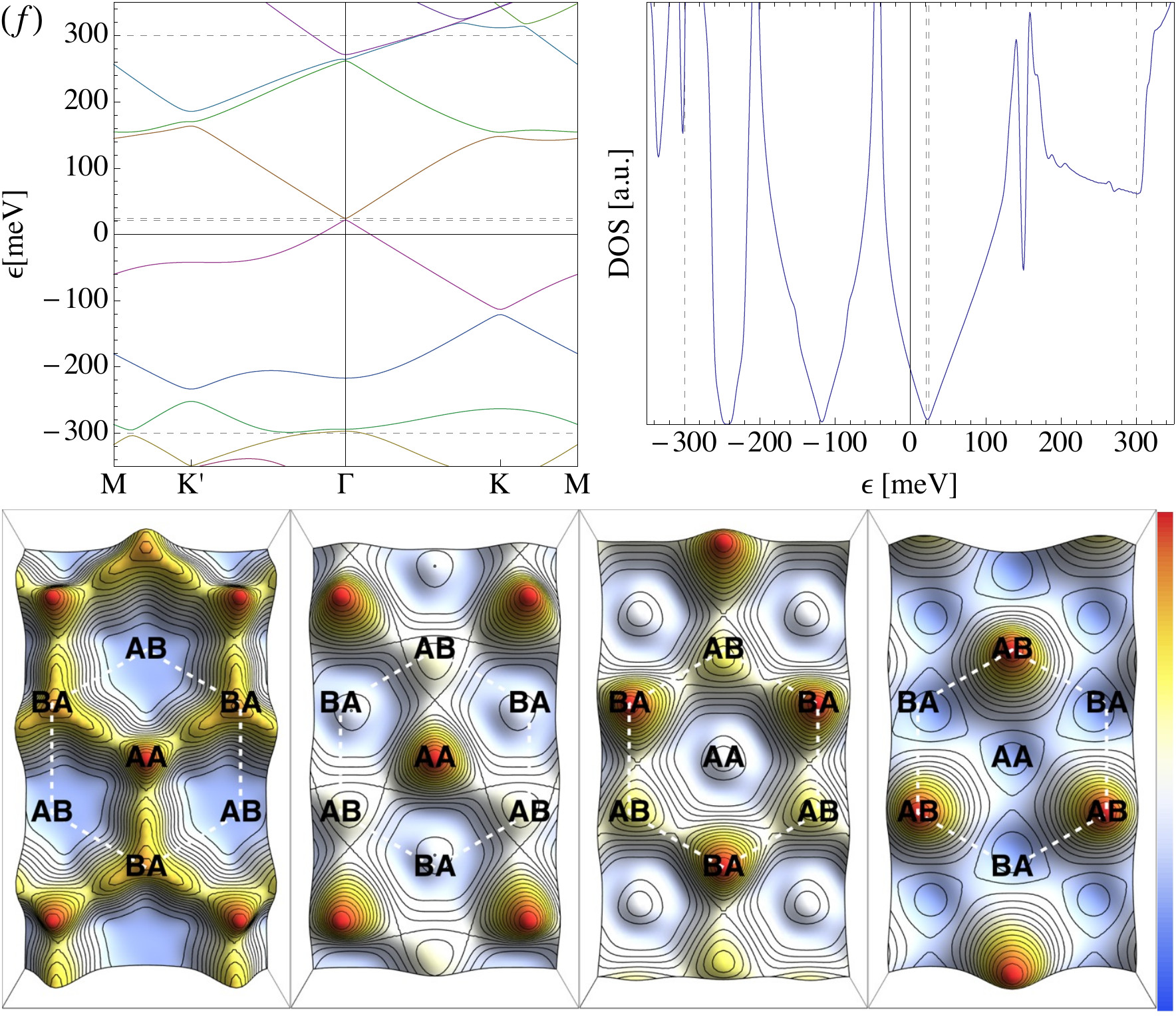}}\\
     $\left(z_u\neq0, t_\perp\neq 0, \lambda_d\neq0, \beta= 0\right)  \rightarrow \Delta\approx 1.8\,\mathrm{meV}$ & & 
     $\left(z_u\neq0, t_\perp\neq0, \lambda_d\neq 0, \beta\neq0\right)  \rightarrow \Delta\approx 1.8\,\mathrm{meV}$
	 \\
     \end{tabular}
     \caption{Electronic structure in the six representative regimes for a model with strains ($z_u\neq 0$), see description for Fig. \ref{fig:ES1}.}
   \label{fig:ES2}
\end{figure*}

The complete effective model has a great deal of structure, derived from the three types of position dependent terms $u_\alpha$ and $\tilde u_\alpha$, plus the uniform mass $\tilde w_3$ and scalar $w_0$ fields. In turn, these terms contain contributions from three competing physical ingredients: (1) the self-energy created by virtual hops into \hBN and back (represented by the $m_\pm\propto t_\perp^2$ terms in all $u_\alpha$, $\tilde u_\alpha$ and $\tilde w_3$), (2) the deformation potential ($\sim \lambda_d$ terms in the scalar potential $u_0$ and $\tilde u_0$), (3) the pseudogauge field ($\sim \beta$ terms in the transverse potential $u_\perp$ and $\tilde u_\perp$). The latter two vanish for $z_u=0$, i.e. in the absence of graphene strains induced by adhesion, while the first remains non-zero. 

In this section we characterise the zero temperature electronic structure of the model in different regimes, through the band structure, density of states (DOS) and local density of states (LDOS) in real space. The goal is to connect different spectral observables to the various physical ingredients in the model, as represented by $z_u$ (strain), $t_\perp$ (interlayer hopping), $\lambda_d$ (deformation potential) and $\beta$ (pseudogauge potential). A complementary analysis, focused mostly on symmetry analysis and on the appearance of secondary naked Dirac points in the DOS, was presented in Ref. \onlinecite{Wallbank:PRB13}, using the $u_\alpha$ and $\tilde u_\alpha$ as control parameters instead.

The results for the band structure, DOS, LDOS and spectral gap, corresponding to all possible non-trivial combinations of parameters $z_u$, $t_\perp$, $\lambda_d$ and $\beta$, is shown in Figs. \ref{fig:ES1} and \ref{fig:ES2}. The simplest case of a strain-free graphene on hBN ($z_u=0$) is shown in Fig. \ref{fig:ES1}(a). The band structure and DOS of the continuum model in this particular case was recently studied in Ref. \onlinecite{Moon:14}. It has an essentially gapless primary Dirac cone (gap $\Delta\approx 0.14$ meV), and exhibits a (slightly gapped) secondary Dirac cone at the K' point in the valence band, possibly naked for certain choice of parameters. This leads to a strong suppression in the DOS at around -170 meV from neutrality, with an associated van Hove singularity. The corresponding LDOS is shown for four energies (dashed lines) also in Fig. \ref{fig:ES1}(a) [contour plots]. It is essentially uniform in space for most energies, particularly at neutrality.

The case with equilibrium deformations but zero pseudogauge and deformation potentials ($\beta=\lambda_d=0$), shown in Fig. \ref{fig:ES1}(b), is mostly similar. The most important difference is the opening of a gap ($\Delta\approx 9.8$ meV for our choice of parameters) in the primary Dirac cone, that is linear in the distortions [$\tilde w_3\sim z_u+\mathcal{O}(z_u^2)$]. This case also exhibits a non-uniform LDOS that is different at either sides of the gap.

The effect of adding the pseudogauge and deformation potentials to the model is shown, in different combinations in Fig. \ref{fig:ES2}. We find that, in general, this tends to enhance the development of secondary, and even tertiary Dirac cones, also in the  conduction band, with the associated van Hove singularities.  Particle-hole symmetry remains broken in general, except for the combination in Fig. \ref{fig:ES2}(a). The primary Dirac cone develops a gap for finite interlayer hopping $t_\perp$, but it can be rather small, since the contribution from pseudogauge and deformation potentials to the gap tend to cancel each other, see Eq. (\ref{gappert2}). The LDOS exhibits a strong spatial and energy dependence in all cases. Fig. \ref{fig:ES2}(f) represent the electronic structure of the full model, including all parameters with the chosen values. We note that the more detailed DOS/LDOS measurements available for this system to date \cite{Yankowitz:NP12} seem to correspond to the spectral profile shown in Fig. \ref{fig:ES1}(b).

\begin{figure}
   \centering
     \begin{tabular}{c}
     \includegraphics[width=0.47\textwidth]{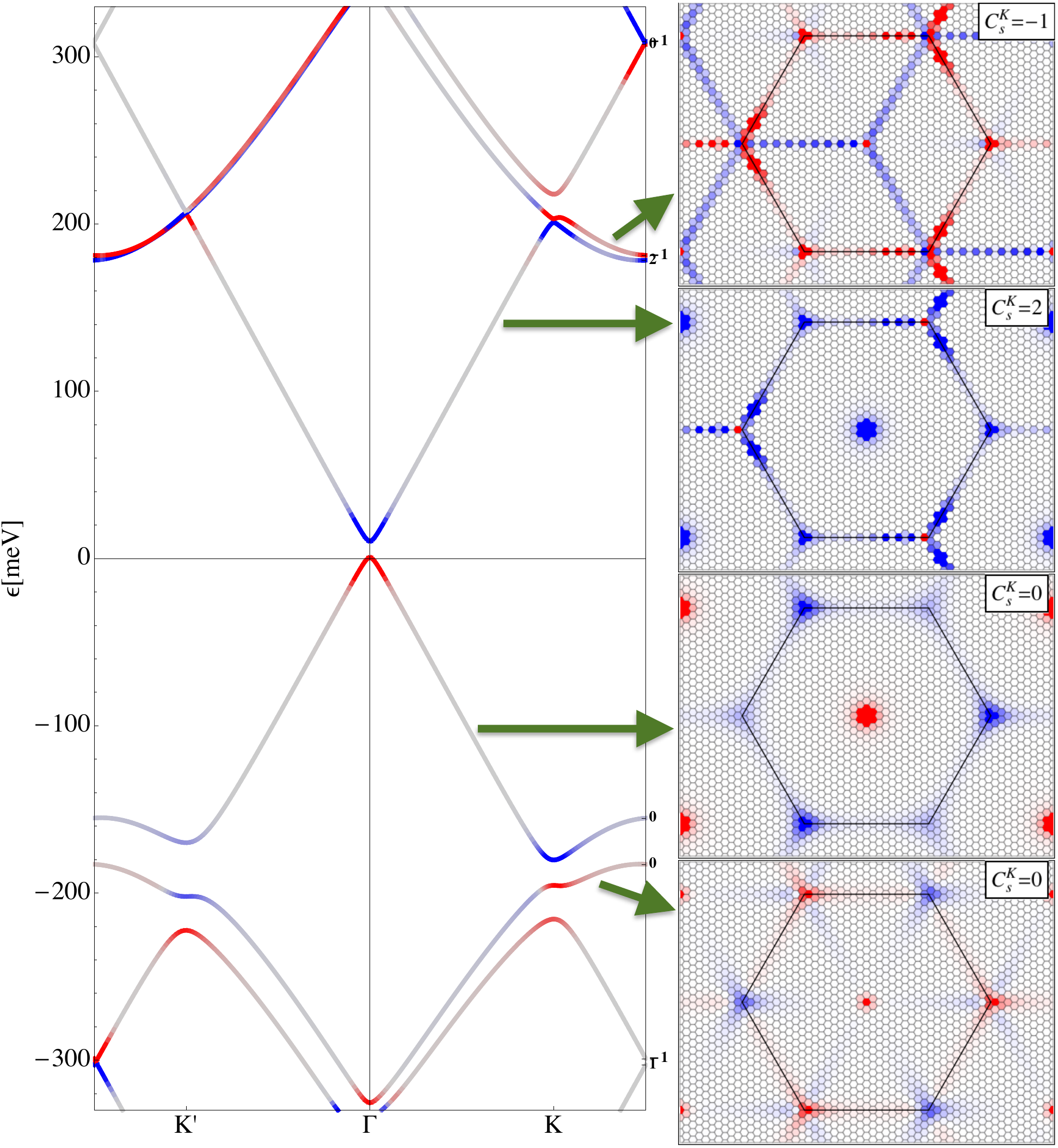} \\
     $\left(z_u\neq 0, t_\perp\neq 0, \lambda_d= 0, \beta=0\right) \rightarrow \Delta\approx 9.8\,\mathrm{meV}$ \\
     \\
     \includegraphics[width=0.47\textwidth]{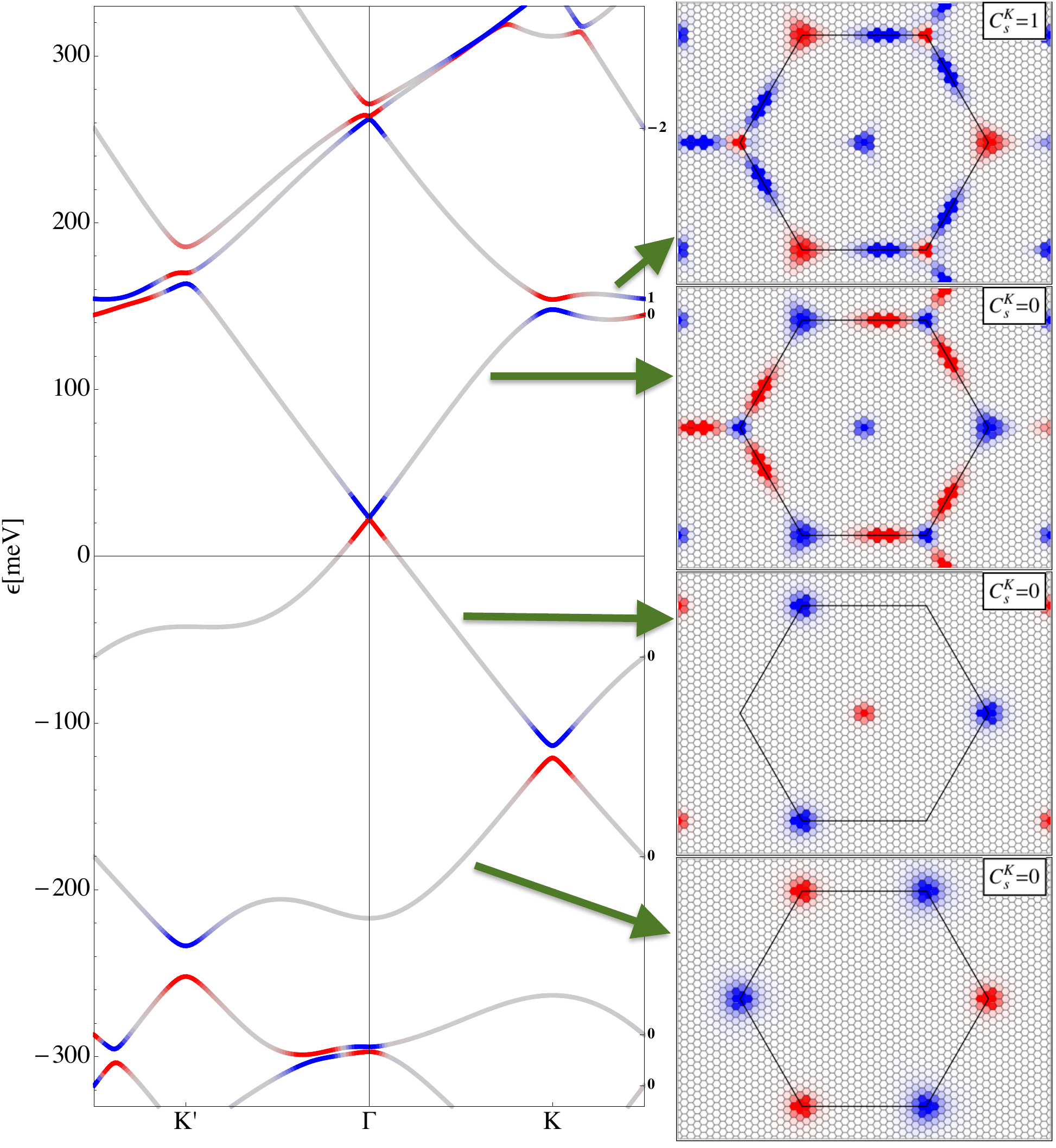} \\
     $\left(z_u\neq0, t_\perp\neq0, \lambda_d\neq 0, \beta\neq0\right)  \rightarrow \Delta\approx 1.8\,\mathrm{meV}$
     \end{tabular}
     \caption{Berry curvature and valley Chern numbers for the lowest subbands of models in Fig. \ref{fig:ES1}(b) (top) and \ref{fig:ES2}(f) (bottom). The four panels on the right depict, for each case, the Berry curvature in the Brillouin zone (black hexagon, with center at the $\Gamma$ point) corresponding to the four subbands closest to zero (blue is positive, red is negative, gray is zero). The Berry curvature is also color coded in the band structure on the left. The valley Chern number $\mathcal{C}_s^{K_\mathrm{gr}}$ for each subband is also shown on the right side of the band structure.}
   \label{fig:Top}
\end{figure}


\section{Band topology}
\label{sec:Top}

Dirac-like models, such as Eq. (\ref{H}), are commonly host to interesting effects associated to non-trivial band topology. The topological charge, or topological invariant, of a subband $s$ in the spectrum is defined as the Chern number $\mathcal{C}_s$, an integer that results from integrating the Berry curvature $\mathcal{F}_s(\vec k)$ over the whole  Brillouin zone (BZ)
\beqa
\mathcal{C}_s&=&\frac{1}{2\pi}\int_\mathrm{BZ} d^2 k \,\mathcal{F}_s(\vec k)\\
\mathcal{F}_s(\vec k) &=&2\im\langle\partial_{k_x}\psi_s(\vec k)|\partial_{k_y}\psi_s(\vec k)\rangle
\eeqa
By time-reversal symmetry the Chern number is $\mathcal{C}_s=0$ for all subbands $s$. This is due to a cancellation of contributions from the neighbourhood of the two valleys $K_\mathrm{gr}$ and $K'_\mathrm{gr}$ in graphene (not to be confused with the $K$, $K'$ points of the superlattice Brillouin zone), $\mathcal{C}_s=\mathcal{C}_s^{K_\mathrm{gr}}+\mathcal{C}_s^{K'_\mathrm{gr}}=0$, where
\beq
\mathcal{C}^{K_\mathrm{gr}}_s=\frac{1}{2\pi}\int_\mathrm{\sim K_\mathrm{gr}} d^2 k \,\mathcal{F}_s(\vec k)=-C^{K'_\mathrm{gr}}_s
\eeq
is integrated close to valley $K$. In valley-symmetric systems, the $\mathcal{C}_s^{K_\mathrm{gr}}$ are themselves quantized, and effectively define a weak topological invariant, under the assumption that any perturbations preserve valley symmetry. Just as a non-zero sum of Chern numbers of completely filled bands are connected to a quantized Hall conductivity $\sigma_{xy}=2e^2/h \sum^{\textrm{filled}}_s \mathcal{C}_s$ (Quantum Hall Effect), non-zero valley Chern numbers imply quantised valley Hall conductivity $\sigma^{K_\mathrm{gr}}_{xy}-\sigma^{K'_\mathrm{gr}}_{xy}=4e^2/h \sum^{\textrm{filled}}_s \mathcal{C}^{K_\mathrm{gr}}_s$ (Quantum Valley Hall Effect) in time-reversal and valley-symmetric systems, such as the one studied here.

Strictly speaking, the valley Chern numbers are only properly defined for subbands that are disconnected from each others. In particular, the Dirac Hamiltonian (as obtained for isolated and undistorted graphene) has no gaps between subbands, and it's valley Chern numbers are ill defined. Upon adding the perturbations described in Eq. (\ref{H}), subbands generically develop avoided crossings, and acquire a well defined topological charge. Ref. \onlinecite{Song:14} studied the valley Chern numbers in the presence of the uniform $\tilde w_3$ and position-dependent $\tilde u_3$ mass terms in different ratios. Here we analyse the band topology for the lowest subbands for the full model in Eq. (\ref{H}). Using the method described in Ref. \onlinecite{Fukui:JPSJ05}, we have computed the Berry curvature and valley Chern number in two relevant cases, the relaxed superlattice with and without deformation and gauge potentials, Figs. \ref{fig:ES2}(f) and \ref{fig:ES1}(b) respectively. The results are shown in Fig. \ref{fig:Top}. The band structure is shaded in red for points with negative, and in blue for positive Berry curvature. The four plots to the right of the band structures show the curvature $\mathcal{F}_s(\vec k)$ in the Brillouin zone (black hexagon) for the four lowest subbands.  We see that for the lowest subbands, particularly the valence subband, the Berry curvature is concentrated at the superlattice $\Gamma$ and $K$, $K'$ points. As argued in Ref. \onlinecite{Song:14}, these have different origins; the former comes from the uniform mass $\tilde w_3$, while the latter depends on position dependent terms $u_\alpha$, $\tilde u_\alpha$. In the case with$\lambda_d=\beta=0$ (Fig. \ref{fig:Top}, top panel), but finite deformations $z_u\neq 0$,  the valence band is trivial ($\mathcal{C}_{-1}^{K_\mathrm{gr}}=0$), whilst the conduction band is non-trivial ($\mathcal{C}_{+1}^{K_\mathrm{gr}}=2$). When deformation and pseudogauge potential are also included (bottom panel), the valence band remains trivial, while the conduction band experiences two band inversions, which make it trivial as well, $\mathcal{C}_{-1}^{K_\mathrm{gr}}=\mathcal{C}_{+1}^{K_\mathrm{gr}}=0$. Both of these cases are in contrast to the results for the corresponding ``commensurate stacking'' regime of Ref. \onlinecite{Song:14}, where the $u_3$, $u_0$ and $\tilde u_0$ terms where neglected, and $\mathcal{C}_{+1}^{K_\mathrm{gr}}=1$ was found. Arguably, the conduction subband at the superlattice $K$, $K'$ points is very close to a band reconnection with the next conduction subband, which makes the value of $\mathcal{C}_{+1}^{K_\mathrm{gr}}$ very sensitive to specific system parameters.

\section{Conclusion}
\label{sec:Conclusion}

We have presented a low energy model that describes the electronic structure of moiré superlattices in graphene deposited on \hBN, when both crystals are crystallographically aligned. The strong spontaneous deformations in this system lead to a strongly modified electronic structure respect to isolated graphene. The electronic model presented here is microscopically derived from a solution to the equilibrium elastic distortions, and compactly integrates a number of previously explored ingredients \cite{Wallbank:PRB13,Jung:14,Moon:14,Song:PRL13a,Song:14} in a single unified analytical description, including moiré-modulated self-energies, and pseudogauge and deformation potentials. We have characterised the resulting electronic structure through the electronic gap, the global and local density of states, the band structure and band topology. 

We have found that the large strains observed in experiments and predicted by theory give rise to a profound transformation of the electronic properties, although the details depend very strongly on the magnitude of specific graphene parameters, particularly the pseudogauge and deformation potentials. Specifically the gap may range from zero to $\Delta\approx 10$ meV for a modulation of $\sim 2.8\%$ in the local expansion, depending on the values considered for $\beta$ and $\lambda_d$. Similarly, the DOS and LDOS exhibit a strong dependence on these parameters. Currently available techniques offer direct measurements of detailed spectral properties in graphene/\hBN superlattices, and may exploit the strong parametric sensitivity of our model to reduce the current uncertainty in the values of these important constants for graphene. More generally, the kind of simple continuum description presented here is a powerful tool to assess the accuracy of our current understanding of the connection between deformations and electronic structure in graphene.

\section{Acknowledgements}
We thank M. Yankowitz and B. Leroy for valuable discussions. We acknowledge support from the Spanish Ministry of Economy (MINECO) through Grant Nos. FIS2011-23713 and PIB2010BZ-00512, the European Research Council Advanced Grant (contract 290846), and the European Commission under the Graphene Flagship, contract CNECT-ICT-604391.

\bibliography{biblio}

\appendix

\section{Microscopic derivation of the low energy effective Hamiltonian}
\label{sec:apHeff}

In this appendix we present a self-contained derivation of the effective low energy model for graphene carriers on an \hBN substrate (modelled without loss of generality as a single layer of \hBN). We consider arbitrary rotation $\theta$ between the crystals, and include the equilibrium strains described in Ref. \onlinecite{San-Jose:14}. The resulting low energy electronic theory is a version of the theory of dos Santos et al. for graphene bilayers, \cite{Santos:PRL07,Santos:PRB12,Moon:14} that is generalised to  incorporates the effects of a strain superlattice. 

\subsection{Interlayer coupling at each valley from the microscopic tight-binding description}

We consider plane waves in the unstrained top and bottom layers, which are given by
\begin{eqnarray}\label{planewaves}
|\psi_{\vec{q},\alpha}^\mathrm{gr}\rangle&=&\frac{1}{\sqrt{N}}\sum_{\vec{n}} e^{i\vec q(\vec r_{\vec n}+\vec \tau_\alpha)}|\vec r_{\vec n}+\vec \tau_\alpha\rangle \\
|\psi_{\vec{q}',\alpha}^\mathrm{\hBN}\rangle&=&\frac{1}{\sqrt{N'}}\sum_{\vec{n}'} e^{i\vec q'(\vec r'_{\vec n'}+\vec \tau'_\alpha)}|\vec r'_{\vec n'}+\vec \tau'_\alpha\rangle \nonumber
\end{eqnarray}
The momenta are measured respect to the $\Gamma$ point of each layer, that is common for both. The sum is performed over all $N^{(')}$ unit cells, centred at $\vec r_{\vec n}$ and $\vec r'_{\vec n}$ for graphene and \hBN, respectively.
These positions can be expressed in terms of the corresponding Bravais vectors $\bm{a}=(\vec a_1,\vec a_2)$ and $\bm{a}'=(\vec a_1',\vec a_2')$ as
\begin{eqnarray*}
\vec r_{\vec n}&=&\bm{a}\vec n\\
\vec r'_{\vec n'}&=&\bm{a}'\vec n'\\
\end{eqnarray*}
for integer $\vec n, \vec n'$, while $\vec \tau_{A,B}=\mp\bm{a}(\frac{1}{6},\frac{1}{6})$ and $\vec \tau'_{A,B}=\mp\bm{a}'(\frac{1}{6},\frac{1}{6})$ represent the position of the two sublattice respect to the center of each unit cell.
The Bravais vectors of the two ayers are geometrically related by $\bm{a}'=\bm{R}\bm{a}$, where
\[
\bm{R}=(1+\delta)\left(
\begin{array}{cc}
\cos\theta & -\sin\theta \\
\sin\theta & \cos\theta
\end{array}
\right),
\]
where $\delta$ is the relative lattice mismatch, and $\theta$ is the rotation angle between the two layers.
Momenta $\vec q$ and $\vec q'$ are defined in the Brillouin zones of each layer, whose basis is given by the rows of $\bm{g}=2\pi\bm{a}^{-1}$ and $\bm{g}'=2\pi\bm{a}'^{-1}$. The two layers form a moiré superlattice, whose minimal conjugate vectors are given by $\bm{G}=\bm{g}-\bm{g}'$ (commensurability subtleties apply \cite{Santos:PRB12}). The Bravais vectors of the moiré superlattice are $\bm{A}=\bm{G}^{-1}/2\pi$, and satisfy $\bm{A}=\bm{a}\bm{N}=\bm{a}'\bm{N}'$, where $\bm{N}=\bm{a}^{-1}(\mathds{1}-\bm{R}^{-1})^{-1}\bm{a}$ and $\bm{N}'=\bm{a}^{-1}(\bm{R}-\mathds{1})^{-1}\bm{a}$.

We now consider that the two layers are a distance $d$ apart. We consider a tight-binding model on the lattice, with interlayer hopping 
\[
T=\sum_{\vec r, \vec r'} t(\vec r'-\vec r){\Psi_{\vec r'}^{\mathrm{\hBN}}}^\dagger \Psi^{\mathrm{gr}}_{\vec r} + \mathrm{h.c.}
\] 
for a certain function $t(\vec r)$ that may be assumed smooth in $\vec r$ on the scale of lattice spacing $a_0$ if $d\gg a_0$. We may express the matrix element of this $T$ in the basis of plane waves above as
\begin{eqnarray}\label{V0}
\langle \psi^\mathrm{\hBN}_{\vec q',\alpha'}| V|\psi^\mathrm{gr}_{\vec q,\alpha}\rangle
&=& \frac{1}{\sqrt{NN'}}\sum_{\vec{n}\vec{n}'} t(\vec r'_{\vec n'}+\vec \tau'_{\alpha'}-\vec r_{\vec n}-\vec \tau_\alpha)\nonumber\\
&&\times e^{i\vec q(\vec r_{\vec n}+\vec \tau_\alpha)}e^{-i\vec q'(\vec r'_{\vec n'}+\vec \tau'_{\alpha'})} 
\end{eqnarray}

We introduce the Fourier transform $\tilde t(\vec p)=\int d\vec r e^{-i\vec p\vec r}t(\vec r)$ of $t(\vec r)$, which is assumed peaked around $|\vec p|=0$. This yields
\begin{eqnarray}\label{V}
\langle \psi^\mathrm{\hBN}_{\vec q',\alpha'}| T|\psi^\mathrm{gr}_{\vec q,\alpha}\rangle
= \int \frac{d^2p}{(2\pi)^2} \tilde t(\vec p)
e^{i(\vec q-\vec p)\vec \tau_{\alpha}}e^{-i(\vec q'-\vec p)\vec \tau'_{\alpha'}}\nonumber\\
\times \frac{1}{\sqrt{NN'}}\sum_{\vec{n}\vec{n}'} 
e^{i(\vec q-\vec p)\vec r_{\vec n}} e^{-i(\vec q'-\vec p)\vec r'_{\vec n'}}
\end{eqnarray}
The last sum imposes a constraint on $\vec p$, $\vec q$ and $\vec q'$ of the form $\sum_{\vec m\vec m'}\delta(\vec q+\vec m\bm{g}-\vec p)\delta(\vec q'+\vec m'\bm{g}'-\vec p)$, where the sum runs over integer $\vec m, \vec m'$. We thus have
\begin{eqnarray*}
\langle \psi^\mathrm{\hBN}_{\vec q',\alpha'}| T|\psi^\mathrm{gr}_{\vec q,\alpha}\rangle
= \sum_{\vec m\vec m'}\tilde t(\vec q+\vec m\bm{g})\delta([\vec q+\vec m\bm{g}]-[\vec q'+\vec m'\bm{g}'])\\
\times e^{i\vec m \bm g\vec \tau_{\alpha}}e^{-i\vec m' \bm g'\vec \tau'_{\alpha'}}
\end{eqnarray*}

%
%
%

Since $\tilde t(\vec p)$ is peaked at the origin, this sum is dominated by the set of $\vec m$ that yield a vector $\vec q+\vec m \bm{g}$ with the smallest modulus. 
We now specialize this model for $\vec q$ close to the K-point of the graphene layer, $\vec q=(\vec g_1-\vec g_2)/3+\vec k=\vec K_\mathrm{gr}+\vec k$, with $|\vec k|\ll|\vec K_\mathrm{gr}|$. In this case, the dominant harmonics correspond to $\vec m\bm g=\{0, -\vec g_1, \vec g_2\}$ (Note these all change sign when considering the neighbourhood of the opposite valley). $\tilde t(\vec p)$ takes the same value for all these three momenta (by symmetry), to lowest order in $\vec k$, 
\[
\tilde t(\vec K_\mathrm{gr})\equiv t_\perp/3.
\]
where the energy scale $t_\perp\approx 0.3$ eV is the interlayer hopping. 
We denote these three values of $\vec m$ as $\vec m_i$, with 
\begin{equation}\label{m}
\vec m_0=(0,0), \hspace{.2 cm} \vec m_{-1}=(-1,0), \hspace{.2 cm} \vec m_2=(0,1).
\end{equation}
Likewise, the $\delta$ constraint ensures that for a $\vec q'$ close to the K-point of \hBN, the allowed values of $\vec m'$ will also be 
$\vec m_i$. Since moreover we must have $\vec q+\vec m\bm{g}=\vec q'+\vec m'\bm{g}'$, we should have $\vec m=\vec m'=\vec m_i$, and $\vec q'-\vec q=\vec m_i(\bm{g}-\bm{g}')=\vec m_i\bm G$.

Therefore, to lowest order in $\vec k\bm A$ around $\vec K_\mathrm{gr}$ and $\vec K_\mathrm{\hBN}$, we have only the following non-zero matrix elements 
\[
T_{i}^{\alpha'\alpha}=\langle\psi_{\vec q+\vec m_i\bm G, \alpha'}^\mathrm{\hBN}|T|\psi_{\vec q\alpha}^\mathrm{gr}\rangle
\]
They read
\begin{eqnarray}\label{VDS}
\bm T_{1}&=&\left(\begin{array}{cc}
1 & 1\\ 1& 1
\end{array}\right) \frac{t_\perp}{3}\\
\bm T_{2}&=&\left(\begin{array}{cc}
1 & e^{-i2\pi/3}\\ e^{i2\pi/3}& 1
\end{array}\right) \frac{t_\perp}{3} \nonumber \\
\bm T_{3}&=&\left(\begin{array}{cc}
1 & e^{i2\pi/3}\\ e^{-i2\pi/3}& 1
\end{array}\right) \frac{t_\perp}{3} \nonumber
\end{eqnarray}
All other matrix elements (including intervalley) vanish. Note that other (less symmetrical) placings of the $e^{\pm i2\pi/3}$ phases above are frequently encountered in the literature. These simply correspond to different choices of $\tau_\alpha$ and $\tau'_{\alpha'}$, or in other words, to a different choice of gauge. 

\subsection{Bilayer Hamiltonian without strains}

Once the relevant matrix elements of the tight binding Hamiltonian are computed, one can write down a low energy continuum model of the whole bilayer. We first write the interlayer hopping matrix in real space
\beq \label{T}
\bm T(\vec r)=e^{i\vec G_{0}\vec r}\bm T_0+e^{i\vec G_{-1}\vec r}\bm T_{-1}+e^{i\vec G_2\vec r}\bm T_2
\eeq
where we denote
\beq
\vec G_i=\vec m_i\bm G, \hspace{0.2 cm}
\vec g_i=\vec m_i\bm g, \hspace{0.2 cm}
\vec g'_i=\vec m_i\bm g'
\eeq
The low energy Dirac Hamiltonian for the graphene layer, with Fermi velocity $v$, reads
\beq
\bm H_\mathrm{gr}(\vec k)=\hbar v \vec k\vec{\bm \sigma}
\eeq
Here, momentum $\vec k$ is measured relative to $\vec K_\mathrm{gr}$. The \hBN layer can be modelled by a similar Hamiltonian, but adding a mass plus scalar term $\bm\Delta_\mathrm{\hBN}$, 
\beq
\bm H_\mathrm{\hBN}(\vec k)=\hbar v' \vec k\vec{\bm \sigma}+\bm\Delta_\mathrm{\hBN}
\eeq
where
\beq
\bm\Delta_\mathrm{\hBN}=\left(\begin{array}{cc}
\epsilon_c & 0 \\
0 & \epsilon_v
\end{array}\right)
\eeq
Here $\epsilon_c\approx 3.34$ eV and $\epsilon_v\approx -1.4$ eV are the conduction and valence band edges of \hBN, relative to graphene's neutrality point. The value of  $v'$ in \hBN is of little consequence in what follows. 

Note that momentum in $\bm H_\mathrm{\hBN}(\vec k)$ is measured relative to $\vec K_\mathrm{\hBN}$, which is different from $\vec K_\mathrm{gr}$ by $\Delta \vec K=\vec K_\mathrm{gr}-\vec K_\mathrm{\hBN}=(\vec G_1-\vec G_2)/3$. Putting this all together we arrive at a model formally identical to that of dos Santos, but that is also valid for a finite lattice mismatch and rotation angle  (given our definition of $\Delta K$)
\beq\label{DSM}
\bm H=\left(\begin{array}{cc}
\bm H_\mathrm{gr}\left(\vec k-\frac{1}{2}\Delta \vec K\right) & \bm T^\dagger(\vec r) \\
\bm T(\vec r) & \bm H_\mathrm{\hBN}\left(\vec k+\frac{1}{2}\Delta \vec K\right)\end{array}\right)
\eeq

An alternative and useful version of the hopping matrix $\bm T(\vec r)$ can be given, \cite{San-Jose:PRL12} that is equivalent to Eq. (\ref{T}),
\beq
\bm T(\vec r)=\left(\begin{array}{cc}
T_0(\vec r) & T_+(\vec r) \\
T_-(\vec r) & T_0(\vec r)
\end{array}\right)
\eeq
Here $T_0(\vec r)=\left(1+e^{-i\vec G_1\vec r}+e^{i\vec G_2\vec r}\right)t_\perp/3$ and $T_\pm(\vec r)=T_0(\vec r\pm\vec r_{BA})$, where here $\vec r_{BA}=(\vec A_1+\vec A_2)/3$ is the center of the BA stacking region in the moir\'e supercell.

\subsection{Bilayer Hamiltonian with strain}

\begin{figure}
   \centering
   \includegraphics[width=0.7\columnwidth]{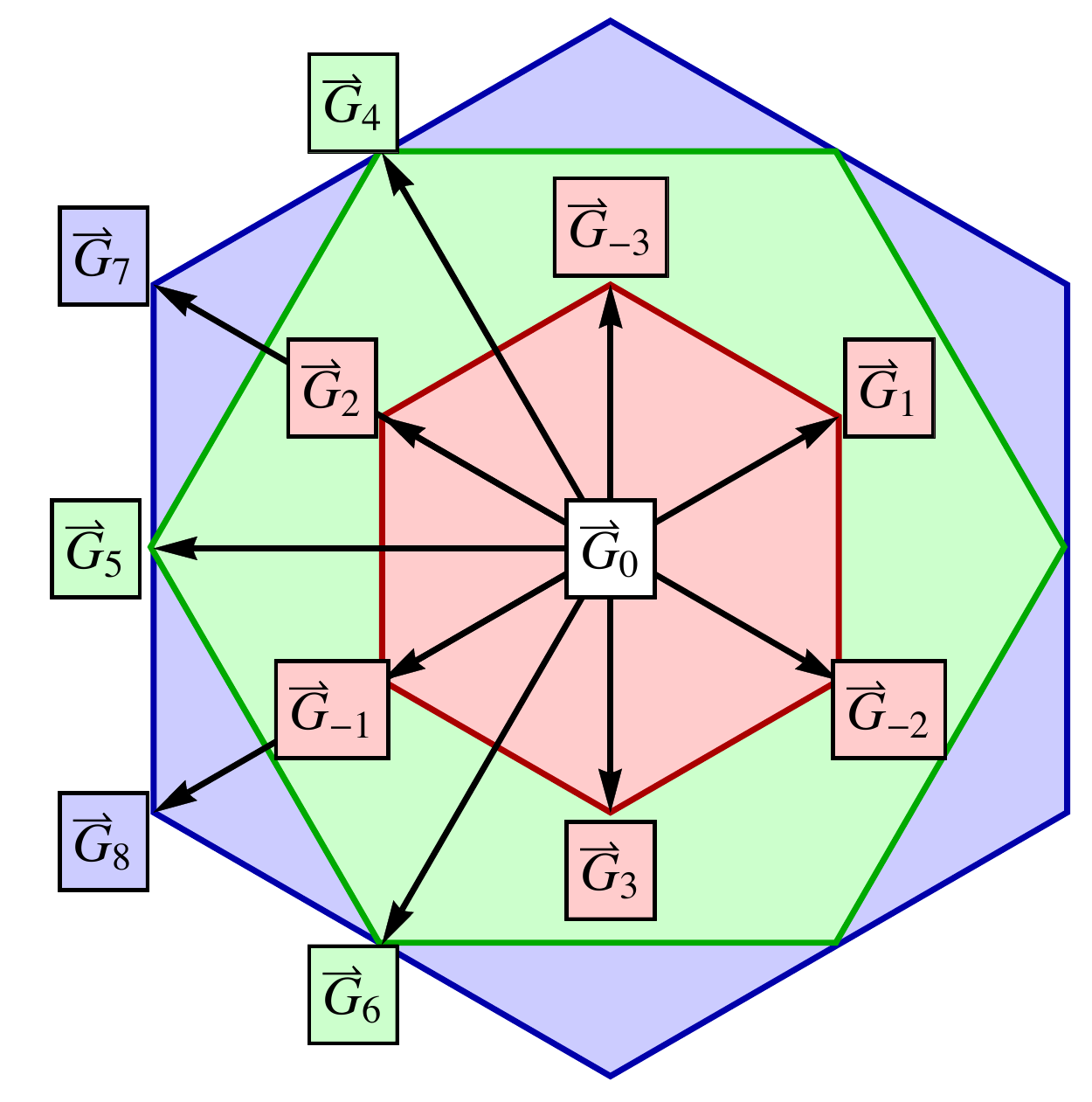}    \caption{Momenta contained in the low energy Hamiltonian $\bm H(\vec r)$ of Eq. (\ref{Hbilayer}). According to their modulus, they are classified within the first star (red), second (green) or third (blue). }
   \label{fig:stars}
\end{figure}

We now consider a distortion field $\vec u(\vec r)$ in the graphene layer. The continuum model Eq. (\ref{DSM}) can cleanly incorporate a small strain field $\vec u(\vec r)$ by modifying the function $T_0(\vec r)$. Alternatively, one can revisit the microscopic derivation to include the distortion. The result is a modified interlayer hopping of the form
\beq
\tilde{T}_0[\vec r,\vec u(\vec r)]=\sum_{j}^{\{0,-1,2\}} e^{i \vec G_j\vec r} e^{-i (\vec K_\mathrm{\hBN}+\vec g'_j)\vec u(\vec r)}
\eeq
where, recall, $\vec K_\mathrm{\hBN}=(\vec g'_1-\vec g'_2)/3$ is the \hBN Dirac point, and $\tilde{T}_\pm[\vec r,\vec u(\vec r)]=\tilde{T}_0[\vec r\pm\vec r_{AB},\vec u(\vec r)]$.

We now assume that the distorion field $\vec u(\vec r)$ is that of Ref. \onlinecite{San-Jose:14}, which is small compared to the lattice spacing, and therefore allows a linear expansion of $\tilde{T}_{\alpha=0,\pm}[\vec r,\vec u(\vec r)]$
\beqa
\tilde T_\alpha\left[\vec r,\vec u(\vec r)\right]&\approx& \tilde T_\alpha\left[\vec r,0\right]+\vec u(\vec r)\left.\partial_{\vec u}\tilde T_\alpha\left[\vec r,\vec u(\vec r)\right]\right|_{\vec u=0} \nonumber\\
&=&\sum_{j}^{\{0,-1,2\}} e^{i \vec G_j\vec r} (1-i (\vec K_\mathrm{\hBN}+\vec g'_j)\vec u(\vec r))\nonumber\\
&=&T_\alpha(\vec r)-i \vec W_\alpha(\vec r)\vec u(\vec r)
\label{Texp}
\eeqa
where we have defined $\vec W_0(\vec r)=\sum_{j}^{\{0,-1,2\}} e^{i \vec G_j\vec r} (\vec K_\mathrm{\hBN}+\vec g'_j)$, and $\vec W_\pm(\vec r)=\vec W_0(\vec r\pm \vec r_{AB})$, so that the interlayer hopping modified by strain reads
\beq \label{tildeT}
\tilde{\bm{T}}(\vec r)=\left(\begin{array}{cc}
T_0(\vec r) & T_+(\vec r) \\
T_-(\vec r) & T_0(\vec r)
\end{array}\right)
-i\left(\begin{array}{cc}
\vec W_0(\vec r)\vec u(\vec r) & \vec W_+(\vec r)\vec u(\vec r) \\
\vec W_-(\vec r)\vec u(\vec r) & \vec W_0(\vec r)\vec u(\vec r)
\end{array}\right)
\eeq
Secondly, we must also incorporate, by minimal substitution, the pseudogauge potential $\vec{\mathcal{A}}(\vec r)$ created by $\vec u(\vec r)$ in the graphene layer, which has the form \cite{Suzuura:PRB02}
\beq
\label{A}
\vec{\mathcal{A}}(\vec r)=\frac{\beta t}{e v}\left(\begin{array}{cc}
\partial_x u_{x}-\partial_y u_{y}\\-\partial_x u_{y}-\partial_y u_{x}
\end{array}\right)\tau_3
\eeq
Lastly, we also include a deformation potential 
\beq
\label{Vd}
\bm V_d(\vec r)=-\lambda_d \sigma_0\left(\partial_x u_{x}+\partial_y u_{y}\right),
\eeq
where the coefficient $\lambda_d$ is an energy somewhere between 5 eV and 30 eV. \cite{Porezag:PRB95,Suzuura:PRB02,Pennington:PRB03,Bruzzone:APL11,Sule:JAP12}
The complete low energy Hamiltonian reads
\begin{widetext}
\beq\label{Hbilayer}
\bm H=\left(\begin{array}{cc}
\bm H_\mathrm{gr}\left[\vec k-\frac{1}{2}\Delta \vec K-\frac{e}{\hbar}\vec{\mathcal{A}}(\vec r)\right] +\bm V_d (\vec r)& \tilde{\bm T}^\dagger(\vec r) \\
\tilde{\bm T}(\vec r) & \bm H_\mathrm{\hBN}\left[\vec k+\frac{1}{2}\Delta \vec K\right]
\end{array}\right)
\eeq
\end{widetext}

Next we consider in more detail the equilibrium displacements $\vec u(\vec r)$ computed in Ref. \onlinecite{San-Jose:14}. These were shown to be a sum of six first-star harmonics
\beq
\vec u(\vec r)=\sum_{j=\pm 1}^{\pm 3} \vec u_j e^{i\vec G_j\vec r}
\eeq
where the complete set of first star momenta $\vec G_j$ are the six red momenta depicted in Fig. \ref{fig:stars}. The equilibrium value of the harmonics $\vec u_j$ read
\beq \label{usol}
\vec u_{j}=iv_j^*\bm{W}_{\vec G_j}^{-1}\vec g'_j
\eeq
where
\beqa
v_{j>0}&=&v_{j<0}^*=(\epsilon_{AA}-\epsilon_{AB})\left(\frac{1}{18}+i\frac{1}{6\sqrt{3}}\right)\\
\bm{W}_{\vec q}&=&(B\det \bm{a})\bm{W}^\parallel_{\vec q}+(\mu\det \bm{a})\bm{W}^\perp_{\vec q}\nonumber\\
\bm{W}^{\parallel}_{\vec q}&=&\left(\begin{array}{cc}
q_x^2 & q_x q_y\\
q_x q_y & q_y^2
\end{array}\right), \hspace{.2 cm}
\bm{W}^{\perp}_{\vec q}=\left(\begin{array}{cc}
q_y^2 & -q_x q_y\\
-q_x q_y & q_x^2
\end{array}\right)\nonumber
\eeqa

The corresponding pseudogauge and deformation potentials of Eqs. (\ref{A}, \ref{Vd}) are thus also first-star, with harmonics
\beqa
\vec{\mathcal{A}}_j&=&
\frac{\beta t}{e v}
\left(\begin{array}{rr}
i\vec G_j\bm \sigma_z\vec u_j\\-i\vec G_j\bm \sigma_x\vec u_j
\end{array}\right)\tau_3 \\ 
\bm V_{d,j}&=&-i\lambda_d\sigma_0 \vec G_j\vec u_j
\eeqa
The resulting $\bm H$ in Eq. (\ref{Hbilayer}) contains harmonics beyond the first start when the above strain is included. Consider in particular the $\vec W(\vec r)\vec u(\vec r)$ corrections in Eq. (\ref{tildeT}), which go up to the third star. The complete set of non-zero harmonics in the model are depicted in Fig. \ref{fig:stars}.

\subsection{Effective monolayer Hamiltonian with strains}

At low energies, electrons are strongly localized on the graphene layer, due to the large gap $\bm \Delta_\mathrm{\hBN}$ in \hBN. It is thus possible to write an effective Hamiltonian that incorporates virtual hopping processes onto the \hBN layer and back in the form of a local self-energy $\bm \Sigma(\vec r)$
\beqa
\bm H_\mathrm{eff}&=&\bm H_\mathrm{gr}\left[\vec k-\frac{e}{\hbar}\vec{\mathcal{A}}(\vec r)\right] +\bm V_d (\vec r)+\bm\Sigma(\vec r) \label{Heff}\\
\bm\Sigma(\vec r)&\approx&-\tilde{\bm T}^\dagger(\vec r)\bm \Delta^{-1}_\mathrm{\hBN}\tilde{\bm T}(\vec r) \label{sigma}
\eeqa
Note that we have gauged away the $\Delta \vec K$ momentum, so that the Dirac point is shifted onto the superlattice $\Gamma$ point. The expression for the local self energy $\bm \Sigma(\vec r)$ is an approximation that is good when $\epsilon_{c,v}$ far exceed the energies under consideration (i.e. up to $\sim 1$ eV around neutrality). In this case, the decoupled Green's function of the \hBN layer can be approximated by $\mathcal{G}(\vec r', \vec r; \omega)=\langle \vec r'|(\omega-\bm H_\mathrm{\hBN})^{-1}|\vec r \rangle\approx -\bm \Delta^{-1}_\mathrm{\hBN}\delta(\vec r-\vec r')$, from which Eq. (\ref{sigma}) follows. Naturally, the number of harmonics in Eq. (\ref{Heff}) is rather large, specifically the set of all pairwise sums of momenta in Fig. \ref{fig:stars}. We have checked, however, that the electronic structure is completely dominated by the first-star harmonics, with the rest contributing only weakly to any spectral or transport observables. It is therefore a good approximation to ignore the latter, and retain only first star harmonics. If we furthermore take into account that the equilibrium deformation $\vec u(\vec r)$ is symmetric under $2\pi/3$ rotations, we can see that $\bm H_\mathrm{eff}$ takes the form given in Eq. (\ref{H}). 

It only remains, therefore, to incorporate the expressions for $\vec u_j$ harmonics from Ref. \onlinecite{San-Jose:14}, evaluate the magnitude of each of the resulting first-star $\bm H_\mathrm{eff}$ harmonics, classify them into even and odd components (i.e. coefficients of $f_1$ and $f_2$, see Eq. [\ref{f12}]), and decompose these into the corresponding Pauli matrices $\sigma_0$, $\sigma_3$, $\sigma_\perp$ and $\sigma_\parallel$, where the latter two are orthogonal and parallel, within SU(2), to the conjugate momentum $\vec G_j$ in question. The algebra is tedious but straightforward, and yields the solution presented in Eqs. (\ref{usolution}) in the particularly relevant case of aligned layers, $\theta=0$.

\section{Corrections to the effective model from inequivalent Carbon-Boron and Carbon-Nitrogen hopping amplitudes} \label{sec:apHop}

In the preceding section we have considered a very general situation, valid for interlayer distance $d$ much greater than the monolayer lattice parameter $a_0$. We have only made one symmetry assumption that need not be exact, namely that the local spatial average of interlayer hopping in AA-stacked region of the moiré and that of the AB- and BA-stacked regions are all the same, and given by $t_\perp$. Deviations from this situation are possible if we assume that hopping amplitudes between Carbon and Boron (which are dominant in AB-regions) and Carbon and Nitrogen (dominant in BA-regions) are different. This difference may be parameterised by a dimensionless quantity $\eta\neq 0$ that modifies the form of $\tilde{\bm T}(\vec r)$ in Eq. (\ref{tildeT}),
\beqa \label{tildeT2}
\tilde{\bm{T}}(\vec r)&=&\left(\begin{array}{cc}
T_0(\vec r) & T_+(\vec r) \\
(1+\eta)T_-(\vec r) & (1+\eta)T_0(\vec r)
\end{array}\right)\\
&&
-i\left(\begin{array}{cc}
\vec W_0(\vec r)\vec u(\vec r) & \vec W_+(\vec r)\vec u(\vec r) \\
(1+\eta)\vec W_-(\vec r)\vec u(\vec r) & (1+\eta)\vec W_0(\vec r)\vec u(\vec r)
\end{array}\right)\nonumber
\eeqa
We must still assume that Carbon atoms in both sublattices of graphene are identical, hence the structure above.
Parameter $\eta$ is exponentially small in the ratio $d/a_0$, and should therefore produce minor corrections. 
One may however consider the corrections for finite $\eta$ to the solution in Eq. (\ref{usolution}). These read
\begin{eqnarray}
\delta w_0&=&\eta(2+\eta)(m_--m_+)\frac{1}{3}\left(1+2z_u+3z_u^2\right) \nonumber\\
\delta \tilde{w}_3&=&0\nonumber\\
\delta u_0&=&\eta(2+\eta)(m_--m_+)\frac{1}{36}\left(1-2z_u-5z_u^2\right)\nonumber\\
\delta \tilde{u}_0&=&\eta(2+\eta)(m_--m_+)\frac{1}{12\sqrt{3}}\left(1-2z_u-5z_u^2\right)\nonumber\\
\delta u_3&=&\eta(2+\eta)(m_--m_+)\frac{1}{12\sqrt{3}}\left(1+2z_u+z_u^2\right)\nonumber\\
\delta \tilde{u}_3&=&-\eta(2+\eta)(m_--m_+)\frac{1}{12}\left(1+2z_u+z_u^2\right)\nonumber\\
\delta u_\perp&=&\eta(2+\eta)(m_--m_+)\frac{1}{18}\left(1+z_u+z_u^2\right)\nonumber\\
\delta \tilde{u}_\perp&=&-\eta(2+\eta)(m_--m_+)\frac{1}{6\sqrt{3}}\left(1+z_u+z_u^2\right)\nonumber\\
\delta u_\parallel&=&\delta \tilde u_\parallel=0
\end{eqnarray}
These expressions should be added to the parameters of Eq. (\ref{usolution}) if $\eta$ is finite. Note, however, that for any finite $\eta$, this does not open a gap to first order, see Eq. (\ref{gap}), since $\delta \tilde w_3=0$. Hence, if the uniform mass $\tilde w_3$ is zero for $\eta=0$ (e.g. for zero strains), it will remains zero for any value of $\eta$.

\end{document}